\documentclass[twocolumn,showpacs,preprintnumbers,footinbib,prd,superscriptaddress,groupedaddress,10pt]{revtex4-1}
\usepackage[utf8]{inputenc}

\makeatletter
\def\l@subsubsection#1#2{}
\def\l@subsubsubsection#1#2{}
\makeatother

\usepackage{graphicx,amssymb,amsmath,amsthm,amsfonts}
\usepackage[usenames]{color}
\usepackage{mathtools}

\usepackage{notes2bib}
\usepackage{adjustbox}

\usepackage{aas_macros}
\usepackage{bm}
\usepackage{dcolumn}
\usepackage{latexsym}
\usepackage{rotating}
\usepackage{longtable}

\usepackage{enumerate}
\usepackage{tensor,multirow}
\usepackage{makecell,array}
\usepackage{url}
\usepackage[linktocpage]{hyperref}

\def\be{\begin{equation}}
\def\ee{\end{equation}}
\def\beq{\begin{eqnarray}}
\def\eeq{\end{eqnarray}}

\def\be{\begin{equation}}
\def\ee{\end{equation}}
	
\newcommand{\bea}{\begin{eqnarray}}
\newcommand{\eea}{\end{eqnarray}}

\begin{document}
\title{Gravitational tuning forks and hierarchical triple systems}
%
%
%
\author{Vitor Cardoso}
\affiliation{CENTRA, Departamento de F\'{\i}sica, Instituto Superior T\'ecnico -- IST, Universidade de Lisboa -- UL,
Avenida Rovisco Pais 1, 1049 Lisboa, Portugal}
\author{Francisco Duque}
\affiliation{CENTRA, Departamento de F\'{\i}sica, Instituto Superior T\'ecnico -- IST, Universidade de Lisboa -- UL,
Avenida Rovisco Pais 1, 1049 Lisboa, Portugal}
\author{Gaurav Khanna}
\affiliation{Department of Physics and Center for Scientific Computing and Visualization Research, University of Massachusetts, Darthmouth, MA 02747\\
Department of Physics, The University of Rhode Island, Kingston, RI 02881}

\begin{abstract} 
We study gravitational wave (GW) emission in the strong-field regime by a hierarchical triple system composed of a binary system placed in the vicinity of a supermassive black hole (SMBH). The LIGO-Virgo collaboration recently reported evidence for coalescences with this dynamical origin. These systems are common in galactic centers and thus are a target for the space-based LISA mission as well as other advanced detectors. Doppler shifts, aberration, lensing and strong amplitude modulations are features present in the GW signal from these systems, built into our framework and with no need for phenomenological patches. We find that the binary can resonantly excite the quasinormal modes of the SMBH, as in the resonant excitation of two tuning forks with matching frequencies. 
The flux of energy crossing the SMBH horizon can be significant, when compared with that from standard extreme-mass-ratio inspirals. Therefore, these triple systems are excellent probes of strong-field physics and of the BH nature of compact objects. 
\end{abstract}
\maketitle

\noindent{\bf{\em Introduction.}}
Since the birth of the gravitational-wave (GW) era in 2015~\cite{PhysRevLett.116.061102}, dozens of GW events have been detected~\cite{Abbott:2020niy}. 
Other detectors will soon join the ground-based network and further improve our ability to measure GWs in the $1-10^3$ Hz frequency range~\cite{Akutsu:2018axf,ET}. The space-based LISA mission will extend detection to the $\sim 10^{-5}-10^{-1}$ Hz window. GWs with these frequencies are emitted in galactic centers by supermassive black holes (SMBHs) and extreme-mass-ratio inspirals (EMRIs), but also by cosmological sources~\cite{Barack:2018yly,Barausse:2020rsu}. The cover of such a broad spectrum will allow us to test General Relativity with unprecedented precision over a wide range of scales, and to answer questions regarding the nature of compact objects, of dark matter and dark energy~\cite{Barack:2018yly,Barausse:2020rsu}.

However, recent results question the validity of the ``standard'' binary system. During its third observation run, the LIGO-Virgo collaboration detected three BH binary coalescences~\cite{LIGOScientific:2020stg,Abbott:2020khf,Abbott:2020tfl,Abbott:2020uma}, unlikely to be composed by two first-generation BHs~\cite{Liu:2020gif, Fragione:2020han}. Instead, their components are thought to be remnants of previous coalescences, forming what is called a ``hierarchical merger''~\cite{Liu:2020gif, Abbott:2020tfl, Fragione:2020han, Martinez:2020lzt,Lu:2020gfh}. Generally, these require the presence of a third body to induce coalescence. 
The Zwicky Transient Facility~\cite{Graham:2019qsw, 2019PASP..131a8002B} reported an electromagnetic counterpart to one of these events, GW195021~\cite{Graham:2020gwr}, consistent with the presence of the BH binary in an active galactic nuclei (AGN)~\cite{Bartos:2016dgn, Stone:2016wzz, 2019ApJ...884L..50M,RevModPhys.82.3121,Ghez_2008}, reinforcing the claim that its components were part of a hierarchical triple system. ``Hierarchical'' here refers to the distinct length scales between the orbit of the BH binary and the one of its center-of-mass (CM) around the third body. Hierarchical triple systems are common in a variety of astrophysical scenarios, such as, globular clusters~\cite{Zevin:2018kzq,Martinez:2020lzt}, AGNs~\cite{Bartos:2016dgn, 10.1093/mnras/stw2260, Chen:2018axp, Toubiana:2020drf}, and other dense stellar environments~\cite{OLeary:2016ayz, 2016MNRAS.463.2109R, Portegies_Zwart_2000}.
Around 90$\%$ of low mass binaries with periods shorter than 3 days are expected to belong to some hierarchical structure~\cite{2006AA...450..681T, Pribulla:2006gk, Robson:2018svj}.

The above motivated recent studies on the dynamics and GW emission in hierarchical triple systems
Kozai-Lidov resonances, in particular, have attracted some attention~\cite{1962AJ.....67..591K, doi:10.1146/annurev-astro-081915-023315, poisson_will_2014}. These describe secular changes in the binary eccentricity and inclination with respect to the orbit described by its CM around the third object. This mechanism triggers periods of high eccentricity ($e \sim 1$) where GW emission increases significantly, potentially inducing coalescence in eccentric orbits detectable by LISA~\cite{Hoang:2019kye, Randall:2019sab, Randall:2019znp, Deme:2020ewx}, which may enter the LIGO-Virgo band still at high eccentricities~\cite{Antonini:2012ad,Antonini_2016, Hoang_2018, Zevin:2018kzq}. Moreover, it can lead to GW bursts at periapsis~\cite{PhysRevD.85.123005,Gupta:2019unn}. A direct integration of the equations of motion 
confirms that GWs from these systems have unique features~\cite{Gupta:2019unn}, which may be detected indirectly via radio observations of binary pulsars~\cite{Suzuki:2020zbg}. There are also attempts at modeling the effects of a third body directly into the waveform. These include Doppler shifts~\cite{10.1093/mnras/stv172, Meiron:2016ipr, Randall:2018lnh, Wong:2019hsq, Han:2018hby}, relativistic beaming effects~\cite{Torres-Orjuela:2018ejx, Torres-Orjuela:2020cly}, gravitational lensing~\cite{Ezquiaga:2020dao, Ezquiaga:2020gdt} and other dynamical effects in triple systems caused by the third-body~\cite{Yu:2020dlm, Bonga:2019ycj, Yang:2019iqa}. 

Studies so far are restricted to the (post-)Newtonian regime and cannot capture strong-field effects. Here, we take a first step towards this direction, and investigate GWs from binaries around SMBHs. Our methods can probe resonant excitation of quasinormal modes (QNMs) in triple systems, and capture for free all of the relativistic effects which have so far been included at a phenomenological level only.
%
%
We adopt units where $c=G=1$.

\begin{figure}	
	\includegraphics[width=7.5cm,keepaspectratio]{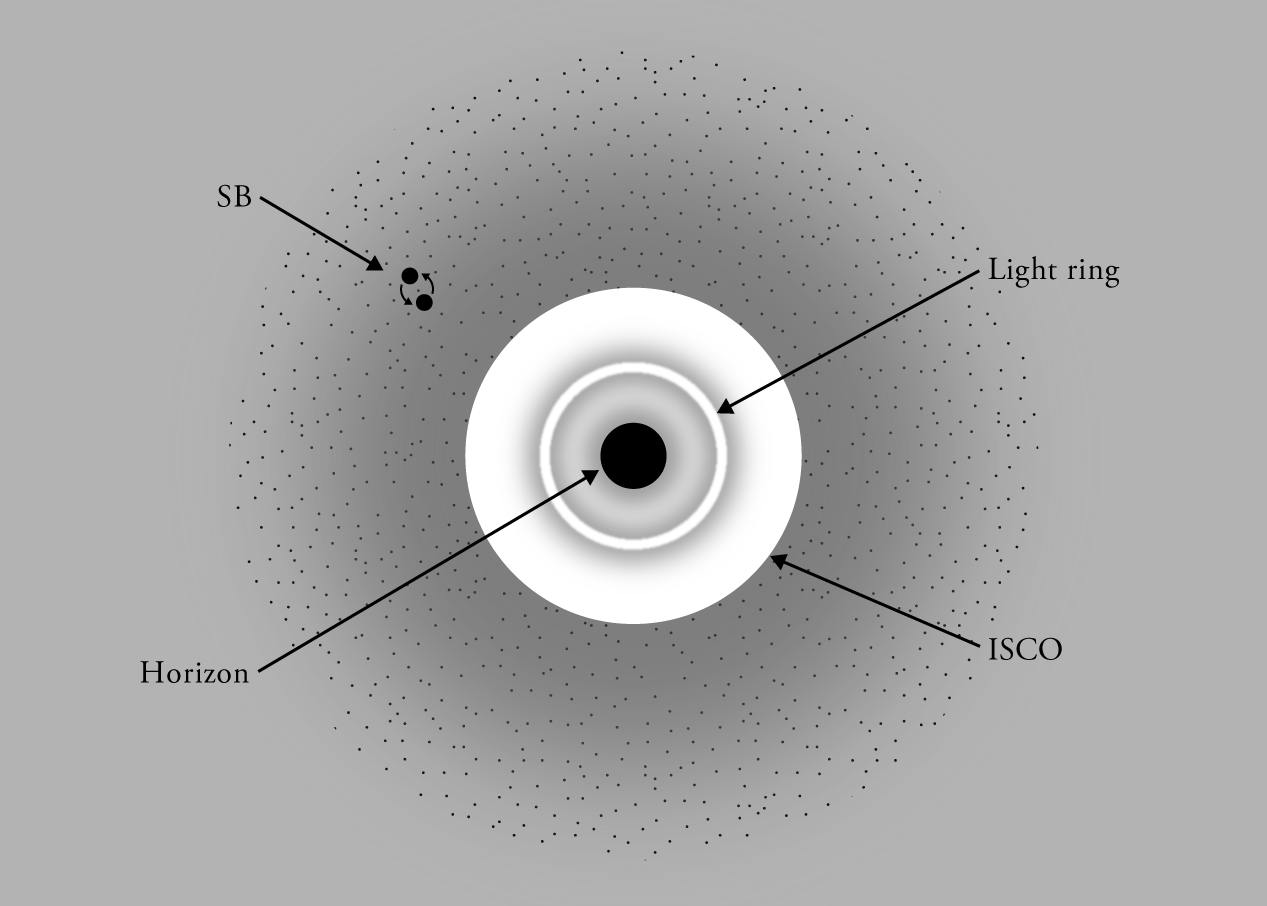} 
	\caption{
	Equatorial slice of a spacetime with a hierarchical triple system, where one component is a central SMBH. We place a small binary (SB) of frequency $\omega_0$ orbiting the SMBH.
	At the innermost stable circular orbit (ISCO), timelike circular motion is marginally stable. High-frequency GWs are (semi-) trapped at the light ring (LR). Such motion is unstable, and can be associated with the ``ringdown'' excited during mergers. Among other effects, here we show that the LR can be excited by tuning $\omega_0$.}\label{fig:anatomy}
\end{figure}

\noindent{\bf{\em Setup: Hierarchical triple systems.}}
We are interested in a setup where a small binary (SB) of compact objects is in the vicinity of a ``large'' BH (larger than all the lengthscales of the SB), as illustrated in Fig.~\ref{fig:anatomy}. The SB is taken to be a small perturbation in a background described by the geometry of the massive BH, which in vacuum must belong to the Kerr family. We use Boyer-Lindquist coordinates $\{t,r,\theta,\varphi\}$~\cite{PhysRevLett.11.237} in our study and define $\Sigma \coloneqq r^2+a^2\cos^2\theta$ and $\Delta \coloneqq r^2-2Mr+a^2$. There is an event horizon at $r_+=M+\sqrt{M^2-a^2}$.
%
%

The SB is modeled as composed of two point particles $\pm$. The SB components also carry each a scalar charge $\alpha$ in our setup, which allows us to study the scalar radiation problem and compare to the more complex gravitational setup. Results for energy fluxes or scalar amplitudes scale in a trivial way with $\alpha$. Since we will only discuss normalized quantities, the actual value of the scalar charge $\alpha$ is not relevant.
If $\tau$ denotes the proper time of each point particle along the world line $z^\mu(\tau)=(t_0(\tau),r_0(\tau),\theta_0(\tau),\varphi_0(\tau))$, the corresponding stress-energy tensor is 
\beq
T^{\mu\nu}(x)^{\pm}&=&m_0^{\pm}\int_{-\infty}^{+\infty}\delta^{(4)}(x-z(\tau))\frac{dz^\mu}{d\tau}\frac{dz^\nu}{d\tau}d\tau\,,
%
\label{eq:StressEnergy}
\eeq
with $\int\int\int\int\delta^{(4)}(x)\sqrt{-g}d^4x\equiv 1$ and $m_0^{\pm}$ is the rest mass of each component of the compact binary.

First-order perturbations on the Kerr spacetime are described by Teukolsky's master equation \cite{Teukolsky:1973ha}  
%
${\cal L}_s\Psi=\Sigma\, \mathcal{T} \label{eq:TeukolskyMaster}$,
%
where ${\cal L}$ is a second-order differential operator, $s$ refers to the ``spin weight'' of the perturbation field (e.g., $s=0,\pm 2$ for scalars and tensors, respectively), and $\mathcal{T}$ is a spin-dependent source term~\cite{Teukolsky:1973ha}.

To compute the source $\mathcal{T}$, we need to prescribe the motion of the SB. We take the CM at $r=R(\tau)$ to either be static at some fixed radius, to describe a timelike equatorial circular orbit around a Kerr BH, or then a simple plunge. 
For the SB inner motion, we take elliptic orbits around the CM, such that
\be
\varphi^{\pm}=\Omega_{\rm CM}t\pm \epsilon_\varphi \sin{\omega_0 t}\, ,\quad \theta^{\pm}=\pi/2 \pm \epsilon_\theta \cos{\omega_0 t} \,,
\ee
where $\epsilon_\theta,\epsilon_\varphi \ll 1$ parametrize the two axis of the ellipse $\delta R_\theta\equiv\epsilon_\theta R$, $\delta R_\varphi\equiv\epsilon_\varphi R$ of the SB and $\Omega_{\text{CM}}$ is the angular velocity of the CM. Note that $\Omega_{\rm CM}$ and $\omega_0$ are coordinate frequencies, while the proper oscillation frequency of the SB, $\omega_0'$, is obtained by a rescaling with the time component of the 4-velocity of the CM, i.e. $\omega_0'=U_{\rm CM}^t\omega_0$.
For concreteness, we focus exclusively on equal-mass binaries, $m_0^\pm=m_0$ and a highly eccentric orbit with $\epsilon_\theta=0$ (we do not see any qualitatively new phenomena in the general case; this particular choice could mimic high-eccentricity binaries driven by Kozai-Lidov resonances). 

A physical relation between $\epsilon_\varphi$ and $\omega_0$ must be imposed. In the SB's rest frame, $\delta R'_\varphi \propto 1/(\omega'_0)^{2/3}$, where the prime refers to \textit{proper} quantities. For SBs on circular geodesics, for example, doing the appropriate rescaling $\omega_0'=U_{\rm CM}^t\omega_0$ and $\delta R_\varphi = \Delta/ \Sigma \,\cdot \delta R'_\varphi $, we find
\begin{eqnarray}
\epsilon_\varphi \propto \frac{\Delta}{\Sigma}\frac{1}{R(U_{\rm CM}^t\omega_0)^{2/3}} \, .  
\end{eqnarray}                                
This relation assumes that the scalar charge $\alpha$ is much smaller than unity and does not affect the motion of the SB in any meaningful way. 

%
We are looking for possible resonances in this triple system, which may happen when the forcing frequency equals natural frequencies of the system. 
%
%
There are three important frequencies in the problem: that of the CM, that of null geodesics on the light ring (LR), and the angular velocity of the BH horizon $\Omega_H=a/(2Mr_+)$~\cite{Bardeen:1972fi}. Close to the BH all are of order $\mathcal{O}(1/M)$, which in fact are also of the order of the QNM frequencies of the central BH~\cite{Berti:2009kk}. To have $M \omega_0 \sim 1$, we need to ensure $\delta R_\varphi / m_0 \sim (M/m_0)^{2/3}$. For a SMBH with $M\sim10^4-10^6 M_\odot$, like Sagittarius A*, and a SB composed by stellar-mass BHs with $m_0 \sim 1-100 M_\odot$, this would correspond to $\delta R/ m_0 \sim 10^2-10^4$. Therefore, the SB can probe the central BH while still well within the inspiral phase of its evolution. Note also that even though Teukolsky's equation assumes very large mass ratios, results in the literature have shown that it is able to reproduce Numerical Relativity for mass ratios of order 10 \cite{Sperhake:2011ik, Rifat:2019ltp}. Hence, our results might extend to the case of an intermediate mass black holes orbiting SMBHs.

\noindent{\bf{\em Numerical implementation.}}
We used two different numerical schemes to solve Teukolsky's equation. One works in the time domain,
and it smooths the pointlike character of the SB constituents~\cite{Krivan_1997,LopezAleman:2003ik,Pazos_valos_2005,Sundararajan:2007jg}. 
The other technique is based on separation of angular variables using spheroidal harmonics~\cite{Berti:2005gp} in the frequency domain, where one can apply standard Green function techniques~\cite{Davis:1971gg,Mino:1997bx,Cardoso:2002ay,Berti:2010ce,Cardoso:2019nis}.
Both approaches are well documented and have been widely tested in the past. 
%
%
Both codes were compared with analytical estimates in the low-frequency regime, obtained using matched asymptotic techniques~\cite{Starobinsky:1973aij,Poisson:1993vp,Cardoso:2019nis}. 
Results from these independent codes are consistent with each other and with analytical estimates.

\noindent{\bf{\em Resonant excitation of QNMs.}}
%
\begin{figure}[t!]
\begin{tabular}{cc}
\includegraphics[scale=0.55]{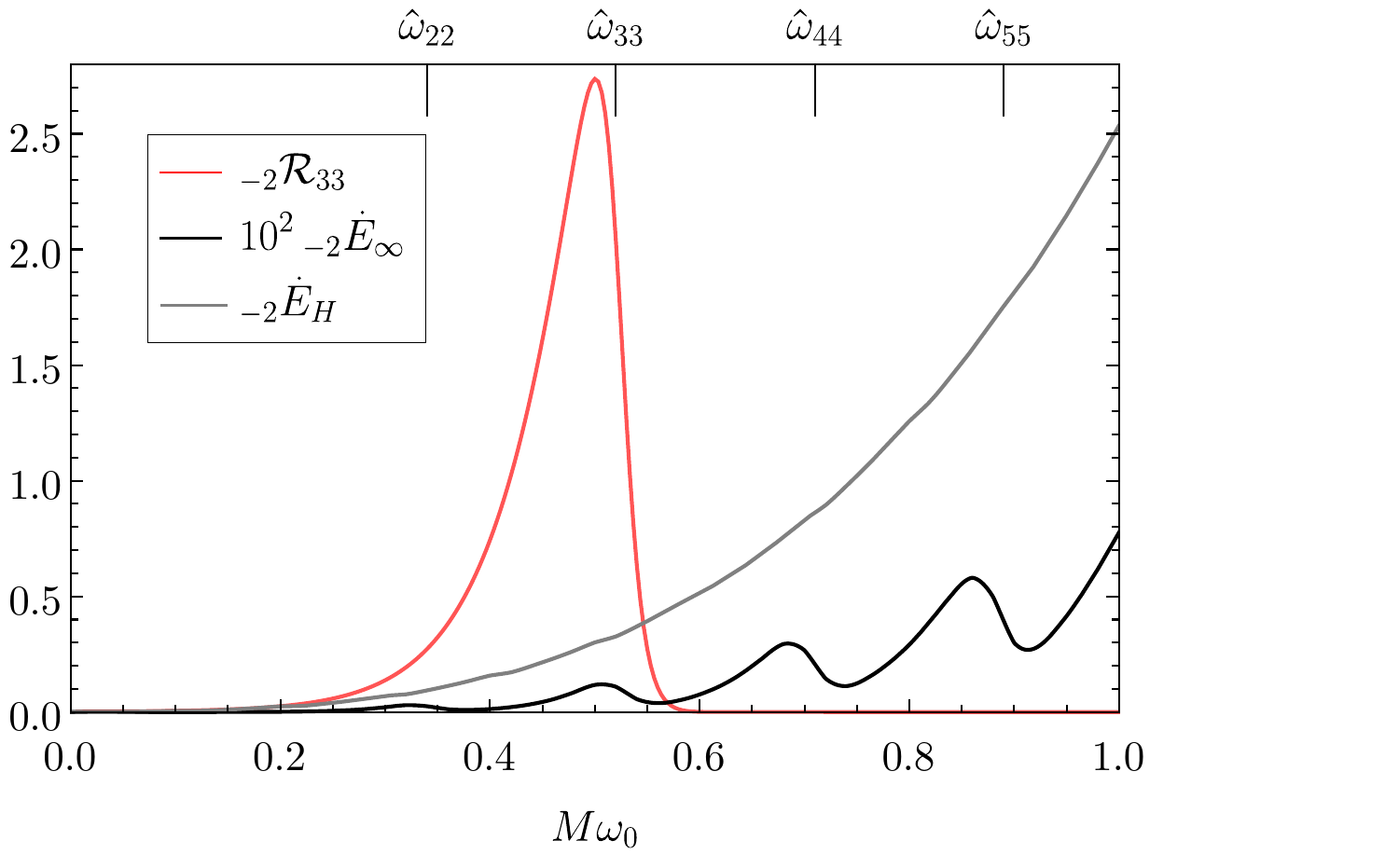}
\end{tabular}
\caption{Energy output when a SB stands at the ISCO of a SMBH of spin $a=0.9M$, as a function of the orbital frequency of the SB components, $\omega_0$. 
The modal energy output, as measured by $_{-2}\mathcal{R}$, peaks at a finite $\omega_0$ extremely well described by the lowest QNM (cf. Table~\ref{tab:MaxFlux}).
Also shown is the flux integrated over all modes: it has a substantial component going down the SMBH horizon, and the total flux at infinity is modulated by QNM contributions.
Here, $\hat{\omega}_{\ell m}\equiv M\omega_{\rm QNM}/2$.
}
\label{fig:Flux} 
\end{figure}
\begin{table}[t!]
\adjustbox{width=\columnwidth}{
\begin{tabular}{lccccccc} \hline\hline
$\ell$ &$s$&$a/M$& $M\omega_\text{QNM}/2$ & $M\omega_{0_\text{LR}}$ &  $M\omega_{0_\text{ISCO}}$ &$_s\mathcal{R}_{\text{LR}}$ &$_s\mathcal{R}_{\text{ISCO}}$\\ 
\hline   
2  &0&0& 0.242 & 0.242 & 0.189  & 4.5  &2.0 \\
2  &-2&0& 0.186 & 0.175 &0.156& 0.6  & 1.5 \\
2  &-2&0.9& 0.335 & 0.332 & 0.319 & 88.0  & 0.8 \\
\hline
3  &0&0 & 0.338 & 0.337 &0.255  & 10.0 &2.5 \\
3  &-2&0& 0.300 & 0.289 &0.250& 2.0  & 2.3	 \\
3  &-2&0.9& 0.522 & 0.520 & 0.500 & 515.8  & 2.7	\\
\hline
4  &0&0& 0.434 & 0.433 &0.317  & 21.6 &3.0 \\
4  &-2&0& 0.405 & 0.395 &0.326& 5.6  & 3.0 \\
4  &-2&0.9& 0.705 & 0.704 & 0.675 & 1896.4 &  5.4	 \\
%
%
\hline\hline
\end{tabular}
}
\caption{Frequency $M\omega_{0\,X}$ which maximizes the energy output of a SB standing at location $X$ close to a SMBH, in a given $(\ell, \ell)$ mode, as measured by the ratio $_s\mathcal{R}$ ($s=0,-2$ for scalar or gravitational perturbations, respectively). The SB CM is {\rm static}, and sitting at the LR or at the ISCO. Notice the excellent agreement with the lowest QNM frequency. The results for orbiting SBs are similar.}
\label{tab:MaxFlux}
\end{table}
We now use the SB as a tuning fork, placing it at some fixed radius, with its CM fixed with respect to distant observers, and letting its frequency $\omega_0$ vary. 
In flat space, this system radiates a (time-averaged) scalar flux in the $\ell\,, m$ mode ($J_\nu(z)$ is a Bessel function of first kind~\cite{NIST:DLMF})
\be
_0\dot{E}_{N \,\ell\, m} = m_0^2\alpha^2\epsilon_\varphi^4\frac{\Gamma\left(\ell+3/2\right)}{64\sqrt{\pi}\, \ell!\, R}\,m^4\,\omega_0\, J_{\ell+1/2}^2(R\,\omega_0)\,,
\ee
and a similar but more cumbersome expression for the Newtonian gravitational-wave flux $_{-2}\dot{E}_{N\,\ell\,m}$. Define an estimate of the SMBH impact through the ratio
\be
_s\mathcal{R}_{\ell\,m} =\, _s\dot{E}_{\ell\,m}/_s\dot{E}_{N\,\ell\,m}\,.
\ee
Our results indicate that at large distances $R$ this ratio tends to unity, as it should on physical grounds.

Figure~\ref{fig:Flux} shows the behavior of $_{-2}\mathcal{R}_{33}$ as the SB frequency $\omega_0$ changes, for an SB sitting at the ISCO of a SMBH.
The behavior is similar for other modes and fields. We observe a peak which we identify as a resonant excitation of the $\ell=m=3$ QNM. As shown in Table~\ref{tab:MaxFlux}, the location of the peak is well described by the lowest QNM frequency~\cite{Berti:2009kk}, for general binary locations. When the SB is placed at the LR, the agreement is excellent (better than 1\% for scalars, and 4\% for GWs for the lowest modes $\ell\,m$ modes). Recall that QNMs can be interpreted as waves marginally trapped in unstable orbits on the photon-sphere~\cite{Cardoso:2008bp}. We therefore arrive at the first result of this paper: a hierarchical triple system behaves as a driven harmonic oscillator~\cite{georgi1993physics}, where the SB is the external harmonic force and the central BH the (damped) oscillator.

This behavior is analogous to the Purcell effect in quantum electrodynamics~\cite{PhysRev.69.37}, describing the enhancement in the spontaneous decay of a quantum emitter inside a cavity, when its frequency matches those of the modes of the field inside the cavity. 
Our results are consistent with recent findings~\cite{PhysRevLett.110.237401}, namely that the spatially independent (i.e. independent of $R$) contribution to the power spectrum in Fig.~\ref{fig:Flux} is described by a Lorentzian curve $\mathcal{R}\propto \omega_{\text{QNM}}^2/(\omega_{\text{QNM}}^2+4 Q^2(\omega_0 - \omega_{\text{QNM}})^2 )$, where 
$Q$ 
is the quality factor of the central BH. Our results are consistent with and extend those of Ref.~\cite{Thornburg:2019ukt}, where resonant excitation of QNMs was observed for EMRIs in eccentric orbits, during passage on the periapsis. The effect is stronger the closer the particle can get to the LR, as also conclude in Ref.~\cite{Price:2015gia}.

As a rule of thumb, the flux peaks at lower frequencies the further the SB is placed from the BH, in agreement with blueshift/redshift corrections. Note that $\mathcal{R}$ smaller than unity does not imply that the system is emitting less energy than expected, since a portion of the radiation falls into the BH. Also, a possible CM orbital motion contributes to a shift in the resonant frequencies by $\pm m \Omega_\text{CM}$, fully consistent with our results. The maximum value of $\mathcal{R}$ in the entire $(R,\omega_0)$ parameter space does not occur precisely at the LR, but close to it. The maximum is attained at locations $R$ closer to the horizon for large $\ell$.
Finally, the magnitude of the resonance grows with $\ell$. For a fixed CM location $R$ and multipole $\ell$ we searched for $\omega_0$ for which $_s\mathcal{R}$ is a maximum $_s\mathcal{R}_{\rm peak}$.
We find an exponential dependence on $\ell$, $_s\mathcal{R}_{\rm peak}\sim a+b\exp(c\cdot \ell)$, at large $\ell$ with $a,\, b,\,c$ constants. 
%

\noindent{\bf{\em Total integrated flux.}}
Ours is a mode decomposition in terms of harmonics of the central BH, thus radiation has support in higher modes as the binary is placed further away from it~\cite{Berti:2005gp, Gualtieri:2008ux}. In general, therefore, the lowest modes will not be dominant and one needs to sum a sufficient amount of modes to understand total fluxes.
%
Already for a SB at the ISCO of a non-rotating BH we find that the GW flux at infinity is comparable to that at the horizon of the SMBH. As seen in Fig.~\ref{fig:Flux},
the effect is more dramatic when spin is included, the flux crossing the horizon can be orders of magnitude larger than that at infinity, even including superradiant modes~\cite{Brito:2015oca}.   
This peculiar aspect is due to the similar length scales of the central BH horizon and the radiation wavelength. GWs are then efficiently absorbed by the BH, in clear contrast with the inspiral phase of an EMRI, whose 
wavelength is much larger than the BH radius. This is our second result: hierarchical triple systems where the SMBH occupies a large fraction of the SB's sky will naturally probe strong field physics, since the fraction of radiation that falls into the SMBH is non-negligible. This will be essential for dynamical evolutions of these systems, particularly when accounting for radiation reaction effects.

For a fixed radius $R$, the field has support on higher $\ell$ modes as the SB is vibrating at higher frequencies $\omega_0$. If the SB is close enough to the BH, it can resonantly excite the QNMs, leading to characteristic peaks 
in the flux at infinity/horizon, as seen in Fig.~\ref{fig:Flux}. These structures correspond to the single multipolar excitations studied in the previous section. 

\noindent{\bf{\em Waveforms: Doppler, aberration \& lensing.}}
%
\begin{figure}[t!]
\begin{tabular}{cc}
\includegraphics[scale=0.41]{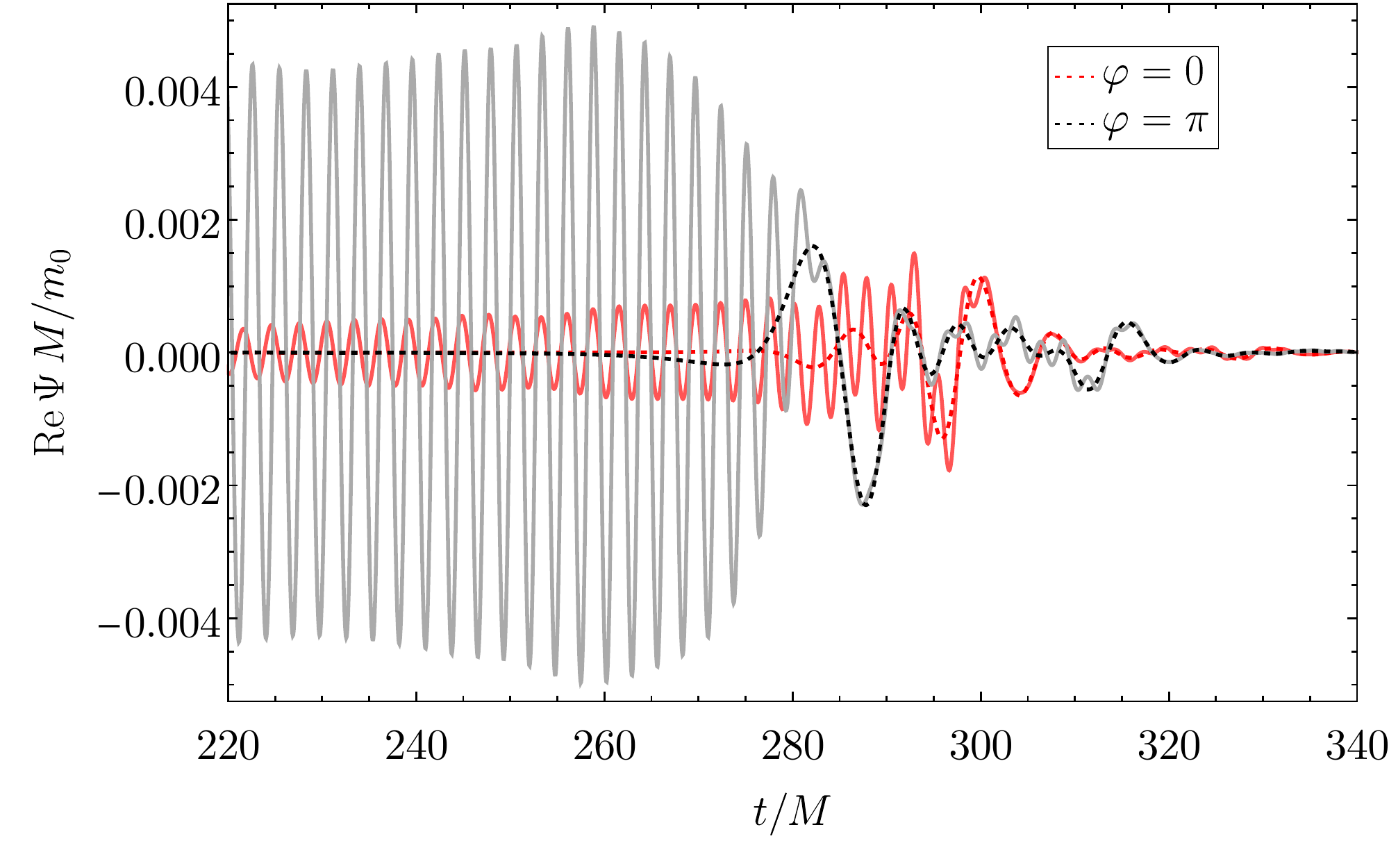}
\end{tabular}
\caption{
Teukolsky function $\Psi$ measured by an (anti)aligned stationary observer at $r= 75M$, for a SB with constant proper frequency $M \omega_0' = 1.0$ radially infalling from $r= 30M$ with zero initial velocity. The dotted lines correspond to the CM contribution to the signal. The SB crosses $r= 10M$ at $t\sim 245M$, the ISCO at $t\sim 263M$ and the LR at $t\sim 278M$.
}
\label{fig:WaveInfall} 
\end{figure}
As a by-product of our methods, we can calculate waveforms from SBs close to SMBHs, which feature interesting relativistic effects.
Figure~\ref{fig:WaveInfall} shows the GW signal produced when a SB, of constant proper frequency $\omega_0'$ falls radially from rest into a non-rotating SMBH. The signal is shown for observers sitting along the merger direction, podal and anti-podal. The observer aligned with the SB sees it moving away, and a GW signal that is progressively redshifted both kinematically and gravitationally (the shifts -- barely visible to the naked eye, are present and agree with expectations). An anti-aligned observer sees a blueshifted signal. As the SB crosses the LR, the radiation it emits is semi-trapped and the signal rings down: the large frequency of the signal is still dictated by the SB, but is now modulated by a low frequency ($\sim 0.19/M$) decay ($\sim e^{-0.1 t}$). The parameters of such decay and low-frequency modulation agree remarkably well with the frequency and damping time of null geodesics at the LR. Imprints of the binary nature of the SB are clearly left on the ringdown stage, that differs visibly from that generated by a point-mass.


%
\begin{figure}[t!]
\begin{tabular}{cc}
\includegraphics[scale=0.55]{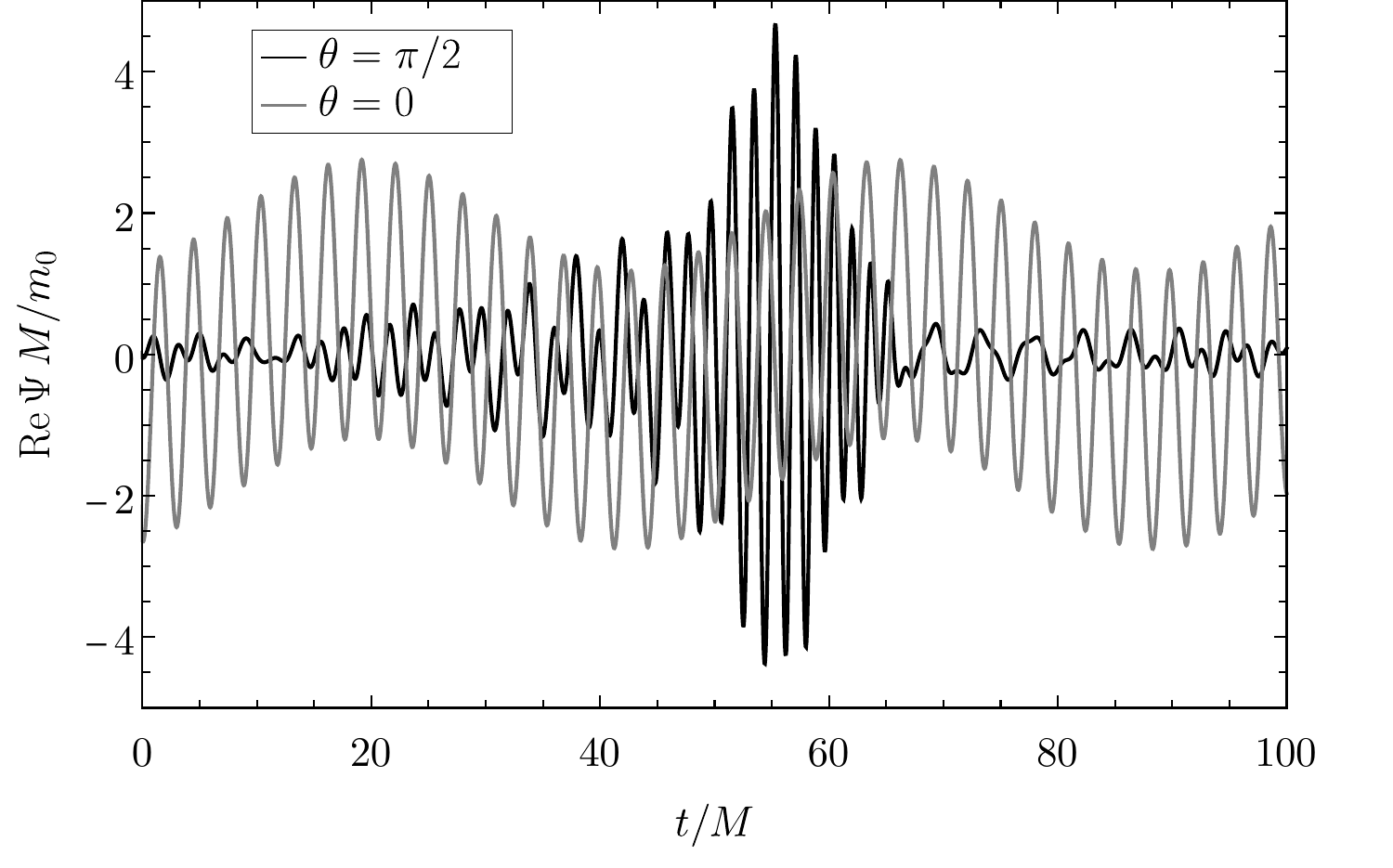}
\end{tabular}
\caption{Teukolsky function $\Psi$ measured by a stationary observer at large distances (either edge- or face-on, $\theta=\pi/2,\,0$ respectively; the face-on signal is multiplied by 100), for a SB around the ISCO of a non-rotating BH (we removed the CM contribution, which just causes 
a low-frequency modulation). The orbital CM period is $T_\text{CM}\approx 93 M$ and at $t=0$ the observer is aligned with the SB. Doppler effect induces frequency shifts, relativistic beaming and gravitational lensing modulations in the amplitude. The maximum blue-shift is well described by $\omega_\text{max}=\omega_0'\Upsilon (\,(\Upsilon+v_\text{CM})/(\Upsilon-v_\text{CM}) )^{1/2}$, with $\Upsilon=\sqrt{1-2M/R}$, $M\omega_0'=1$ the proper SB frequency and $v_\text{CM}$ is the CM velocity~\cite{1972ApJ...173L.137C, 10.1093/mnras/stv172}.}
\label{fig:WaveCirc} 
\end{figure}
Finally, Fig.~\ref{fig:WaveCirc} shows the GW measured by stationary observers at large distances, for a SB on circular motion at the ISCO of a non-rotating BH. These are signals calculated from first-principles. We removed the (linear) CM contribution, which only induces a low-frequency modulation. 
Observers on the equatorial plane see gravitational and Doppler-induced frequency shifts, consistent with analytical predictions~\cite{1972ApJ...173L.137C, 10.1093/mnras/stv172} when the CM is moving towards the observer. The amplitude of the wave can vary by orders of magnitude because of relativistic beaming~\cite{Torres-Orjuela:2018ejx,Torres-Orjuela:2020cly, Gupta:2019unn} and gravitational lensing~\cite{Ezquiaga:2020spg, Ezquiaga:2020gdt}. The former focuses the radiation along the direction of motion, and is significant for fast CM motion.
The maximum amplitude does not occur precisely when the SB is moving towards the observer ($t\sim 70 M $ in Fig.~\ref{fig:WaveCirc}) but slightly before, when the SB is still behind the BH with respect to the observer. This is due to lensing by the central BH, which distorts the path taken by GWs and concentrates radiation on certain directions, amplifying the signal~\cite{Nambu:2015aea,Nambu:2019sqn}. This effect is more relevant for larger frequencies, when the radiation wavelength is much smaller than the BH radius.
On the other hand, observers facing the plane of motion ``face-on'' ($\theta = 0$) do not measure such modulations, since the motion of the CM is now transverse. The only feature is a modulation in amplitude coming from the CM motion (at second order), which has also been reported in Post-Newtonian studies of triple systems~\cite{Gupta:2019unn}. 

\noindent{\bf{\em Discussion.}}
We show that a stellar-mass binary system (or any other radiator) in the vicinity of a SMBH is an excellent probe of strong gravity.
Under special circumstances, which require a fine tuning of the system,
the binary can resonantly excite the modes of the SMBH, offering a unique opportunity to probe the Kerr geometry and the presence of horizons in the cosmos. Even if this fine tuning is not present, the comparable order of magnitude between the SB's radiation wavelength and the SMBH horizon radius leads to an enhancement of energy absorption by the SMBH for any frequency.

Such classes of hierarchical triple systems are abundant in AGNs, and thus our results have implications for GW astronomy, in particular for LISA which is specially designed to detect GWs originated in galactic centers~\cite{Barausse:2020rsu}. While quantifying a detectability rate for the resonances we described goes beyond the scope of this work, we can estimate if a SB can get close enough before being tidally disrupted due to the Hills mechanism~\cite{1988Natur.331..687H, Addison:2015bpa, Suzuki:2020vfw}. This occurs if the tidal forces induced by the BH overcome the binary's self gravity, which happens at a radius $R_t \sim 2\delta R  \left(M/2m_0\right)^{1/3}$. The SB frequency will be related to its separation by the Kepler's law $\omega_0 \sim \sqrt{2 m_0/\delta R^3}$. We thus find $R_t\lesssim 1/(M\omega_0)^{2/3} M$. Already for $M \omega_0 = 0.2$, we find that tidal disruption happens at $R_t \sim 5.84 M$, smaller than the ISCO of a Schwarzschild BH. Thus, SBs very close to a central BH and oscillating at relevant frequencies of the system have astrophysical interest. This is supported by more sophisticated numerical works \cite{Brown:2018gar}.
We neglected spin-spin effects in the motion of the SB. The corrections are proportional to $\sigma=q J/m_0^2$, with $J$ the angular momentum of the SB~\cite{Jefremov:2015gza}. Again using Kepler's law, one finds that corrections to the motion scale like $\sigma\propto q^{2/3}$, which are extremely small for the systems we consider. 

A follow-up to our work is to study the capacity of 
GW detectors to distinguish between these systems and isolated binaries. In particular, it is important to quantify the systematic errors incurred in parameter estimations from a signal originated in a hierarchical triple, using GW templates for isolated binaries. 
Moreover, it is important to extend our study to other motions. An interesting case is a SB describing a high-eccentricity orbit around a spinning SMBH. Such eccentric orbits can be formed naturally in non-trivial environments~\cite{Cardoso:2020iji}. In these orbits, the SB gets closer to the LR, which 
enhances the resonant excitation of the SMBH \cite{Thornburg:2019ukt} and may lead to manifestations of superradiance~\cite{Brito:2015oca}.
Another interesting triple system is a pair same-sized BHs 
and a third lighter compact object orbiting around them. These spacetimes have been shown to have global properties not present in isolated BHs (e.g. global QNMs)~\cite{Bernard:2019nkv, Ikeda:2020xvt} and our results suggest that the lighter object can excite these global modes.

\noindent{\bf{\em Acknowledgments.}}
We thank Ana Carvalho for producing some of the figures in this work.
We are grateful to Béatrice Bonga, Emanuele Berti, Hirotada Okawa and Paolo Pani for useful comments and suggestions.
We thank UMass Darthmouth and Waseda University for warm hospitality while this work was being finalized. F.D. is indebted to Nur Rifat and Asia Haque for help provided during his stay in UMass Darthmouth.
V.C. acknowledges financial support provided under the European Union's H2020 ERC 
Consolidator Grant ``Matter and strong-field gravity: New frontiers in Einstein's 
theory'' grant agreement no. MaGRaTh--646597.
F.D. acknowledges financial support provided by FCT/Portugal through grant No. SFRH/BD/143657/2019. 
G.K. would like to acknowledge support from the National Science Foundation (NSF) under awards PHY-2106755 and DMS-1912716. 
This project has received funding from the European Union's Horizon 2020 research and innovation 
programme under the Marie Sklodowska-Curie grant agreement No 101007855.
We thank FCT for financial support through Project~No.~UIDB/00099/2020.
We acknowledge financial support provided by FCT/Portugal through grants PTDC/MAT-APL/30043/2017 and PTDC/FIS-AST/7002/2020.
The authors would like to acknowledge networking support by the GWverse COST Action 
CA16104, ``Black holes, gravitational waves and fundamental physics.''
%

\bibliographystyle{h-physrev4}
\bibliography{references}

\end{document}